\documentclass[prb,showpacs,preprint,superscriptaddress]{revtex4-1}
\usepackage{amssymb}
\usepackage{graphicx}
\usepackage{bm}

\usepackage{epsfig}
\usepackage{amsfonts}

\begin{document}

\title{Near-threshold properties of the electronic density of layered
quantum-dots}

\author{Alejandro Ferr\'on}
\email{aferron@conicet.gov.ar}
\affiliation{Instituto de Modelado e Innovaci\'on Tecnol\'ogica
(CONICET-UNNE), Avenida Libertad 5400, W3404AAS Corrientes, 
Argentina}

\author{Pablo Serra}
\email{serra@famaf.unc.edu.ar}
\affiliation{Facultad de Matem\'atica, Astronom\'{\i}a y F\'{\i}sica,
Universidad Nacional de C\'ordoba and IFEG-CONICET, Ciudad Universitaria,
X5016LAE C\'ordoba, Argentina}

\author{Omar Osenda}
\email{osenda@famaf.unc.edu.ar}
\affiliation{Facultad de Matem\'atica, Astronom\'{\i}a y F\'{\i}sica,
Universidad Nacional de C\'ordoba and IFEG-CONICET, Ciudad Universitaria,
X5016LAE C\'ordoba, Argentina}

\begin{abstract}
We present a way to manipulate  an electron trapped in a
layered quantum dot based on near-threshold properties of one-body
potentials. Firstly, we show that  potentials with a simple global parameter
allows the manipulation of the wave function changing its spatial localization.
This
phenomenon seems to be fairly
general and could be
implemented using current quantum-dot quantum wells technologies and materials
if a proper layered quantum dot is designed.
So, we propose a model layered quantum dot that consists of a spherical core
surrounded by successive layers of different materials. The number of layers and
the constituent material are chosen to highlight the near-threshold properties.
The manipulation of the spatial localization of
the 
electron in a layered quantum dot 
results consistent with actual experimental parameters.
\end{abstract}
\date{\today}

\maketitle

\section{Introduction}
The tailoring of particular quantum states has become an usual task in quantum
information processing \cite{Brunner2011}. Semiconductor quantum dots (QDs) are
ideally suited to storage quantum information through its eigenstates, 
since electrons in QDs can store quantum phase coherence for very long periods 
of time \cite{Takahashi2011}. In addition the amount of entanglement that a 
multi-electron QD can keep in storage could be useful to implement quantum 
information tasks, but there exist very few examples of ab initio calculations 
that attempt to quantify it in the literature 
\cite{FOS,pots,pos,bili,dami1,dami2}. 
The dots can be designed in a host of ways to meet the specific requirements 
of the quantum information task in sight, by choosing its shape, size or
materials \cite{Reimann2002}. 

On the other hand, the advances in semiconductor technology allow the
preparation of more complex structures than the simple quantum well.
Between the  more complex structures, the quantum-dot
quantum wells structures \cite{Schoos1994} and multiple quantum rings
\cite{Mano2005}
have been extensively studied.  The quantum-dot quantum wells (QDQW's)
structures are
multi-layered quantum dots,  composed of two semiconductor materials, the
one with the smaller bulk band gap is sandwiched between a core and an outer
shell of the material with larger bulk band gap. Because of its properties, the
QDQW's have been demonstrated
to form an efficient gain medium for nanocrystal-based
lasers \cite{Xu2005}, and its electronic structure has been obtained from
first-principles calculations \cite{Schrier2006}. It is remarkable that the
effective mass approximation (EMA) seems to predict fairly well the behavior 
of the electronic density \cite{Schrier2006,Berezovsky2005,Meier2005} 
when compared with first-principles calculations.

The quantum ring structures are an ideal playground to study many subtle
quantum phenomena as the Aharonov-Bohm effect  \cite{Bayer2003} which leads to
the
presence of persistent currents \cite{Mailly1993}. This persistent currents
have been measured even for one electron states \cite{Kleemans2007}. The
quantum rings are formed by only one
semiconductor material, and the fabrication of multiple concentric 
quantum rings
(up to five) can be achieved with high quality and reliability
\cite{Somaschini2009}.  

Despite all the advantages that quantum dots present as implementations of
physical qubits,  the  confined electrons  interact with thousands of
spin  nuclei through the hyperfine interaction. This leads, inevitably, to
decoherence. To solve the decoherence problem it has been proposed that the
physical qubit should be implemented by two electrons confined in a double
quantum dot \cite{Petta2005}. The coupling of the two quantum dots depends on
the spatial extent of the one-electron wave functions, in particular it
determines the strength of the exchange interaction between the two electrons.
The exchange coupling in double quantum dots has been studied extensively, in
particular how it can be tuned using electric fields \cite{Kwa2009}, magnetic
fields \cite{Szafran2004}, or the effect of the confinement of the double
quantum dot in a quantum wire \cite{Zhang2008}. So, the ability to manipulate
the spatial extent of the one electron wave function in an isolated QD, can be
decisive when dealing with the states of a double quantum dot. 

In addition, the recent advances in materials manipulation at the nanoscale
allow an accurate shape and size control of the QD that offers
the possibility of tailoring the energy spectrum to produce desirable
optical transitions. These tasks are useful for the development of
optical devices with tunable emission or transmission properties. Optical
properties of artificial molecules and atoms are subject of great interest
because of their technological implementations. In spherical QD, the optical
properties such as the dipole transition, the oscillator strength and
the photoionization cross section have been studied theoretically by different
authors\cite{o1,o2,o3,o4,o5,o6}.

In this work we present a way to manipulate the wave function of an electron 
trapped in a quantum dot whose energy is near the continuous threshold. The 
phenomena associated seem to be fairly general and could be implemented using 
current quantum-dot quantum wells technologies and materials.
In particular we show that the phenomenon is present in a layered quantum
dot model whose parameters are consistent with actual experimental values.

The paper is organized as follows. In Sec. II we introduce a
simple  model  whose near-threshold behavior shows how the
localization of the ground state electronic density could be handled easily. In
Sec. III we  apply the findings of Section II to design a
 QDQW-like 
 model that is both consistent with actual experimental values and able
to exploit the near-threshold phenomenon.
 Here
we perform calculations for the ground state and the first excited $p$ 
state, and we
analyze the optical properties of the model. Finally, Sec. IV contains the
conclusions with a
discussion of the most relevant points of our findings.

\section{One Electron oscillating potential}

In this section we report results for the ground state electronic density 
of one electron in an oscillating short range potential. The near
threshold properties will be exploited in the next Section
to design
quantum dot structures. We choose an smooth and analytical potential to
emphasize that the behavior of the electronic density is fairly general. 
The one particle Hamiltonian is given by
\begin{equation}\label{hamiltoniano}
H = -\frac{\hbar^2}{2m} \nabla^2 + W(r),
\end{equation}
where

\begin{equation}\label{potential2}
 W(r) = - W_0 e^{-\gamma r} \sin(\omega r) ,
\end{equation}

\noindent where $W_0$ is the strength of the potential and 
$\gamma$ and $\omega$ are positive constants.  
Clearly, the potential well range 
is given by $\gamma$ and the number of  spherical layers and their
width  are related 
to $\omega$. The successive spherical layers are similar to a succession of
potential wells and barriers. 

It is known that for fixed
values of
$\gamma$ and $\omega$, there is a critical value $W_0^{(c)}$ such that the 
Hamiltonian Eq. (\ref{hamiltoniano}) supports at least one bound state only if
$W_0 > W_0^{(c)}$ \cite{kais03}.
In the following, we will study the near-threshold properties of the
ground state electronic density varying $W_0$ as a global parameter, for $W_0
\gtrsim W_0^{(c)}$.

\begin{figure}[floatfix]
\begin{center}
\psfig{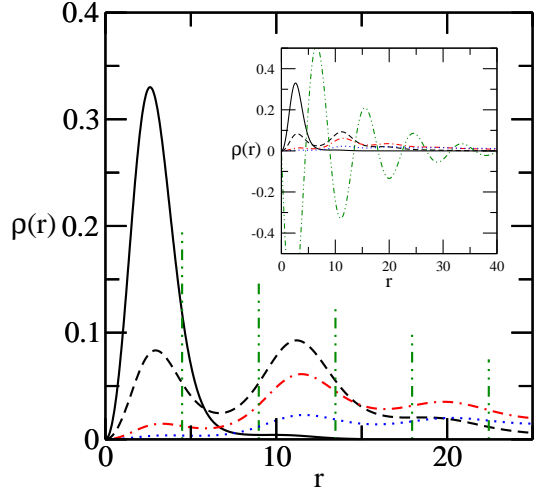}
\end{center}
\caption{\label{electronicden}(color on-line) Electronic density 
as a function
of $r$ for the ground state Hamiltonian (Eq. (\ref{hamiltoniano})). The dark 
green vertical dashed doted lines show the zeros of the 
potential defined in Eq. (\ref{potential2}). The black continuous
line shows the electronic density for $W_0=0.3\, a.u$, the black dashed
line shows the electronic density for $W_0=0.2\, a.u$, red 
dashed doted line shows calculation for $W_0=0.16\, a.u$ and
 blue  doted line  shows the electronic density for $W_0=0.14
\, a.u$. The inset shows the same electronic densities and the potential in a
different scale. 
The  dark green  dashed doted curve
corresponds to the  potential, which is plotted in $W_0$ units.}
\end{figure}

The ground state energy and wave function of Hamiltonian Eq. 
(\ref{hamiltoniano}) were
obtained numerically, using the fourth-order Runge-Kutta method for the
potential in Eq.~\ref{potential2}.
The  electronic density, $\rho(r)$, 
is given by
\begin{equation}
 \rho(r) = \int \, |\Psi(\mathbf{r})|^2 r^2 \, d\Omega,
\end{equation}
where $\Psi$ is the one-electron wave function.
We are interested
in the probability of finding the electron in the n-th well. 
Figure~\ref{electronicden}
shows the electronic density $\rho(r)$ for different values of $W_0$ . By
changing the value of $W_0$ it 
is possible to select in which potential well the maximum value of the 
electronic density is located. The particular value of $W_0$ that locates the 
maximum of the electronic density in, say, the third well depends on the 
actual values of $\gamma$ and $\omega$.

\begin{figure}[floatfix]
\begin{center}
\psfig{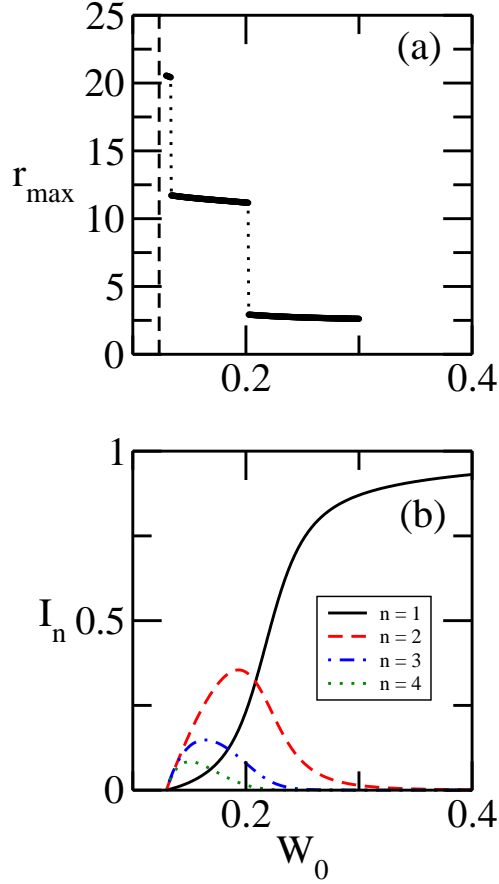}
\end{center}
\caption{\label{rmax}(color on-line) 
(a) Position where the electronic density achieves its maximum value 
($r_{max}$) as a function of $W_0$. The vertical dashed line shows the
critical ionization value, $W_0^{(c)}$.
(b) The well occupation probability, $I_n$, as a function of $W_0$. For large 
values of $W_0$ the ground state wave function is localized around the origin, 
so only $I_1$ is appreciable. For smaller values of $W_0$, $I_1$ drops to 
zero and the successive wells became occupied. Near the threshold many wells 
are occupied but $I_n\rightarrow 0, \;
\forall n$.}
\end{figure}

In Figure~\ref{rmax}a we can observe the position  where the electronic 
density achieves its maximum value ($r_{max}$). We can check that the value
$r_{max}$ is rather stable as a function of $W_0$. Moreover, Figure~\ref{rmax}a 
shows that changing $W_0$ allows selecting in which well the electronic density
will have its maximum value. It is worth to mention that this effect is a
near-threshold phenomenon different from those
reported previously in the literature  for
electrons localized in complex nanostructures \cite{Schoos1994,Szafran2004}
, since
in our example the maximum is not necessarily located in the deepest well.

The maximum in the electronic density as a function
of the strength potential shows an interesting phenomenon. We can observe
how this maximum jumps from
a well to the next well when the strength potential is continuously
varied. It is important to note that we have a situation 
where the maximum is not located in the deepest well. In order to
clarify this phenomenon we define the ``well occupation probability'',

\begin{equation}\label{ine}
I_n =\int_{2 (n-1) \pi/\omega}^{(2 n -1) \pi/\omega}  \rho(r) dr
\end{equation}

\noindent that allows us to evaluate how much of the electronic density
is in the n-th well. In figure~\ref{rmax}b we show the numerical calculations
of $I_n$ for different values of $n$. The intersection of the black solid line
($I_1$) and the red dashed 
line ($I_2$), of Figure~\ref{rmax}b,
give us the value of $W_0$ where a qualitatively change in the electronic
density occurs. For this value of $W_0$ we have the first jump. The 
second jump occurs for the value of $W_0$ where the red dashed line and the 
blue dashed dotted line intersect.

\begin{figure}[floatfix]
\begin{center}
\psfig{figure=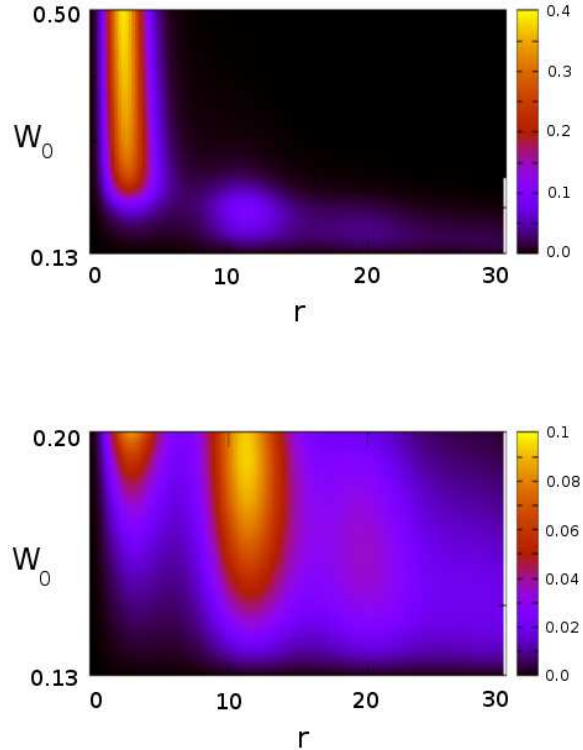,width=8cm}
\end{center}
\caption{\label{map}(color on-line)
The upper panel shows the contour map of the electronic density as a function
of the coordinate $r$ and the strength potential $W_0$ for the potential 
defined in Eq. ({\ref{potential2}}). A zoom of this figure is shown in the
lower panel. The figures show 
how the electronic density's higher peak moves as $W_0$  decreases.
Note that for certain values of $W_0$ the electronic density
has one, two or three peaks.}
\end{figure}

Finally in Figure~\ref{map} we plot the Contour map of the electronic 
density as a function of the coordinate $r$ and the strength potential $W_0$ for
the potential defined in Eq. ({\ref{potential2}}).
Here we can appreciate clearly the  effect reported in this work. For
larger values of $W_0$ ($W_0>0.25$) the electronic density presents just one 
peak located in the first well. When we decrease the value of
$W_0$ we start to see a second (lower) peak located in the second well, but
for values of $W_0\simeq 0.2$ the second peak is higher than the first peak
and the maximum is not located in the deepest well. If we continue decreasing
$W_0$ the peak in the first well vanishes and we start to appreciate
a new peak in the third well.

\section{One electron QDQW-like model}

The ability to produce easily distinguishable quantum states in nanodevices is
of fundamental importance because of its potential technological applications,
in this sense we show that the spatial extent of the ground state electronic
density can be noticeably changed, without changing the spatial extent of the
nanostructure.The ground states  could be distinguished by
current imaging  techniques \cite{Patane2010}. 

On the following we show how our findings could be relevant in actual physical
implementations. The synthesis of layered quantum dots is a well known
technique, see Reference \cite{Schoos1994} and references therein. Moreover,
the EMA approximation appropriately describes the electronic structure of
nanostructures formed by layers of CdS and HgS. Since the modulation of a
global parameter like our $W_0$ is not easily achievable, and  that the
potential well in CdS/HgS compounds is given by the conduction band offset
between the two materials, the only parameters that can be varied with some
ease by experimentalist are the width  and the
materials of the layers. Another difference when dealing with hetero
nanostructures with the simple model Hamiltonian in
Eq. (\ref{hamiltoniano}) results from the different effective mass
characteristic of each material.

Here we consider structures with two
wells made of a  HgS layer, separated by one CdS step and with a central CdS
core. The step-like potential can be written  

\begin{equation}
\label{potrc}
V(r) = \left\lbrace 
\begin{array}{lll}
V_0,& \quad & r < r_c; (CdS), \\
0, & &  r_c\leq r<r_{1}; (HgS) ,\\
V_0,& & r_1\leq r <r_2; (CdS),\\
0,& & r_2\leq r <r_3; (HgS) ,\\
V_0,& \quad & r \geq r_3 ; (CdS).
\end{array}
\right.
\end{equation}

\noindent where the potential well depth is $V_0=1.35$ $eV$, which corresponds
to the band offset between CdS and HgS, while the effective masses are $0.2m_e$
and $0.036m_e$ respectively \cite{Bryant1995}. 
The width of the inner and outer HgS layers were fixed equal to 2 and 1.5 $nm$,
respectively,
while the width of 
 CdS layer that separates them was fixed to 2 $nm$. 
The width of the central 
CdS core is going to be varied in order to show the phenomenon presented
in previous section. An important difference with 
$W_0$ is that the width of the layers can be varied easily in the laboratory. 


\subsection{Ground state electronic density}
Here we evaluate the ground state electronic density for the 
layered QD given by Equation (\ref{potrc}), as we did it in the previous
section.

\begin{figure}[ht]
\begin{center}
\psfig{figure=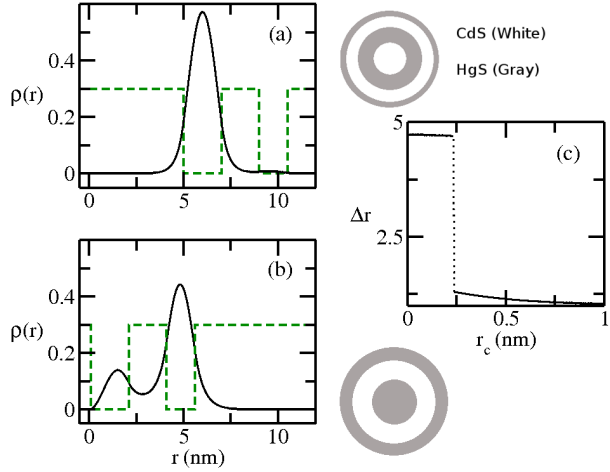,width=8cm}
\end{center}
\caption{\label{expe}(color on-line) 
The electronic density for two different layered quantum
dots. The electronic density in panel (a) corresponds to a
CdS/HgS/CdS/HgS/CdS structure ($r_c=2\,nm$), while panel (b) corresponds to a
HgS/CdS/HgS/CdS one ($r_c=0\,nm$). The structures are shown as ring patterns 
(after the last HgS layer we consider that we have a very long CdS barrier). 
The panel (c) shows the behavior of the position where the electronic 
densities achieves its maximum value ($\Delta r=r_{max}-r_c$), measured with 
respect to the inner radius of the first potential well as a function of 
the core width $r_c$. For large enough 
values of the CdS core the electronic density maximum lies in the first 
potential well, as shown in the panel (a). When the radius of the CdS core 
approaches $0.5\,nm$ the maximum of the electronic density jumps to the second 
potential well, as shown by the abrupt change in $\Delta r$ in panel (c). The
blue dashed line in (a) and (b) shows the step like potential.}
\end{figure}

As Figure~\ref{expe} shows, by changing the width of the CdS core the
electronic density maximum can be located in the potential well of election.
Besides, the position of the maximum, as a function of the core width, shows
the
same behavior that the electronic density of the model Hamiltonian, compare the
Figure~\ref{expe}c with Figure~\ref{rmax}a.

The stability of the electronic density's maximum can be used when dealing with
coupled structures. The fabrication of nanodevices is subjected to many errors
so, at least in principle, it could be very useful to have nano-structures
whose properties are not excessively sensitive to the actual fabrication
parameters. 

\subsection{Optical Properties}
Optical properties are related to the transitions between different states 
of the quantum system. In this section we need to evaluate the ground state
and some of the excited states of the one electron system. As we 
mentioned before, the phenomenon under investigation is a near threshold
property of the system, because of that, the calculation of the excited
states requires some effort.

\begin{figure}[ht]
\begin{center}
\psfig{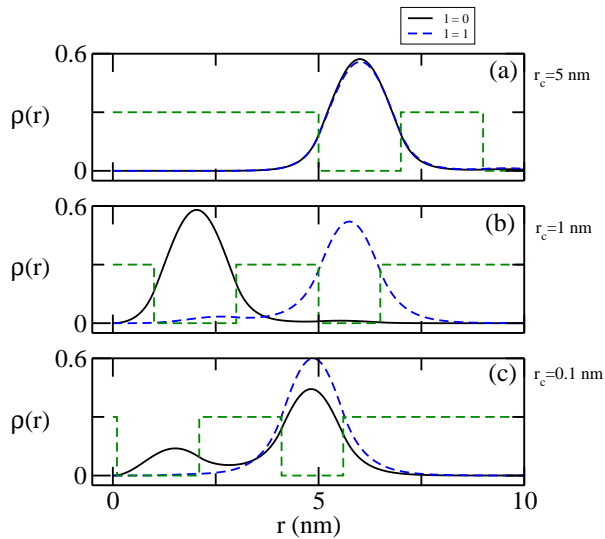}
\end{center}
\caption{\label{expe2}(color on-line) The electronic density for three 
different CdS/HgS/CdS/HgS/CdS layered quantum dots. The electronic density in 
panel (a) corresponds to $r_c=5\,nm$, while panel (b) 
$r_c=1\,nm$ and panel (c) $r_c=0.1\,nm$. 
The dark green dashed line in (a), (b) and (c) shows the step like potential.}
\end{figure}

In Figure~\ref{expe2} we show the electronic densities for the ground 
state and the lower excited state with angular momentum $l=1$. 
In
Figure~\ref{expe2} (a)
we can see that the maximum of both electronic densities are located in 
the first potential well. As we decrease the width of the central core ($r_c$)
we observe that the maximum value of the excited state electronic density
jumps to the second potential well while the ground state electronic density
remains almost unchanged (Figure~\ref{expe2} (b)). Finally, for small
values of the central CdS core we obtain the picture presented before.
In Figure~\ref{expe2} (c) we can see that the maximum of both electronic 
densities ($l=0$ and $l=1$) are located in the second potential well. It is
reasonable to 
assume that this issue must be reflected in the transition rates between these
states, and therefore in the optical properties of the one electron 
QDQW.

It is worth to mention that the overlap between the ground state and the excited $p$ state 
varies from almost one (see Figure \ref{expe2}a) to a very small value 
(see Figure \ref{expe2}b), and back (see Figure \ref{expe2}c), without
changing very much the size of the device. This feature can not be achieved with 
single well QD without changing its size over one or two magnitude orders since 
for very large QD the eigenfunctions become very delocalized, resulting in a large 
overlap between them.

In order to analyze the optical properties of the one electron layered
quantum-dots we are going to investigate the electronic dipole-allowed 
transitions. With this purpose we calculate the Oscillator Strength $P_{fi}$ 
that is a dimensionless quantity that can be evaluated using the
expression\cite{bassani}

\begin{equation}
P_{fi}=C\Delta E_{fi} |M_{fi}|^2 ,
\label{osf}
\end{equation}

\noindent where $M_{fi}$ is the dipole transition matrix element
between the lower ($i$) state and the upper state ($f$),
$\Delta E_{fi}=E_f-E_i$ and $C$ is a normalization
constant. Using the
Thomas-Reiche-Kuhn sum rule

\begin{equation}
\sum_{n} P_{0n}+\int_0^\infty P_{0E}dE=1 ,
\end{equation}
 
\noindent the normalization constant can be calculated as

\begin{equation}
C=\left\langle 0\left|\frac{1}{m^{\star}(r)}\right|0\right\rangle^{-1} .
\end{equation}

\begin{figure}[ht]
\begin{center}
\psfig{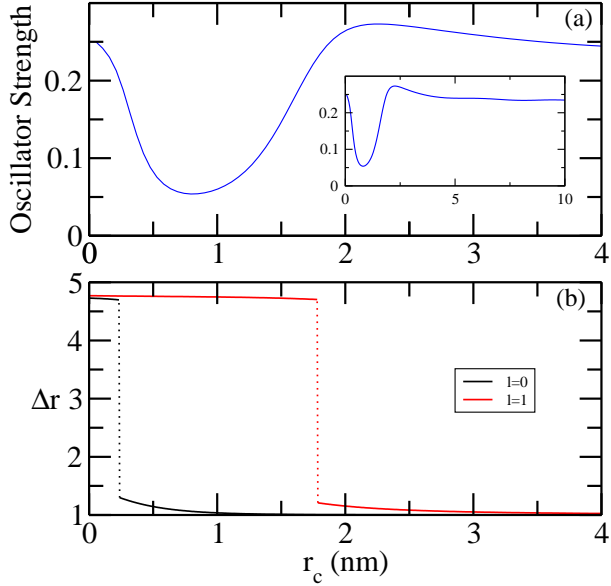}
\end{center}
\caption{\label{os}(color on-line) Panel (a) shows the oscillator strength
as a function of the core width $r_c$
calculated using Equation (\ref{osf}) for the ground and lower excited state
with $l=1$. Panel (b) shows the behavior of the position where the electronic
densities achieves its maximum value ($\Delta r=r_{max}-r_c$), measured with
respect to the inner radius of the first potential well as a function of
the core width $r_c$. For large enough
values of the CdS core the electronic density maximums lie in the first
potential well. When the radius of the CdS core
approaches $1.8\,nm$ ($0.23\,nm$) the maximum of the excited sate 
(ground state) electronic density jumps to the second potential well, as shown 
by the abrupt change in $\Delta r$.}
\end{figure}

In Figure~\ref{os}a we can observe the oscillator strength for the transitions
between the ground state and the lower excited state with $l=1$ as 
a function of the core width. In the lower panel,  Figure~\ref{os}b
we show, as in Figure~\ref{expe}(c), the behavior of the position where the 
electronic densities (ground and excited states) achieves its maximum 
value ($\Delta r=r_{max}-r_c$), measured with respect to the inner radius 
of the first potential well as a function of the core width $r_c$. 
For large enough values of the CdS core the electronic density maximums lie 
in the first potential well. As the radius of the CdS core decreases the 
maximums of the electronic densities jumps to the second
potential well ($r_c=1.8 nm$ for the excited state density and $r_c=0.23 nm$ 
for the ground state density), as shown by the abrupt change in $\Delta r$. 
As we can see in the inset of Figure~\ref{os}a the oscillator strength 
rapidly decrease for large values of the width of the central CdS core
and then reaches to a limit value. The behavior for small values 
of $r_c$ is rather more complicated, and as was pointed out before, 
the phenomenon becomes evident. It is clear from Figure~\ref{os}a and 
Figure~\ref{os}b 
that when the maximum values of the electronic densities are located
in different potential wells, the oscillator strength presents a dramatic
decrease.

\section{Conclusions} 

We have shown a complete analysis of the ground state of one electron
in a spherical potential described in Eq. (\ref{potential2}) and
(\ref{potrc}). We find a effect that, to the best of our knowledge, has not 
been reported in the literature. For the potential defined in (\ref{potrc})
we analyzed the ground and the lower excited state with $l=1$. The
effect mentioned before is also present in the excited state. We finished
the present work with the study of the optical properties of the
one electron layered quantum-dots. Here we find that the effect is
 detected by the oscillator strength, that is a very important physical
quantity in the study of the optical properties, and is related to the 
electronic dipole-allowed transitions. Moreover, we can conclude that
we can have a great control of the optical properties of these nanostructures
in a simple manner using the well known techniques of synthesis of layered
quantum dots. A small change in the width of the central core can 
produce dramatic changes in the optical properties of QDQW.

This phenomenon can be experimentally
observed with the actual semiconductor technology. One possible setback for
the  observation of the  phenomenon discussed in this work comes from the
low binding energies associated to  near-threshold phenomena. Given
the fairly simple dependence of the reported phenomenon on the potential
characteristics, we believe that its observation is possible and feasible. For
the model analyzed in this paper, the availability of materials with larger band
offsets could render the phenomenon more pronounced and with larger eigenenergies.

The behavior of the
electronic density in one and two dimensional systems with a potential
equivalent to
Eq. (\ref{potential2}) is qualitatively the same. Anyway, since the 
near-threshold behavior in two dimension is not exactly the same 
that the observed
in three dimensions, it is possible that the jumping of the electronic density
between 2-D potential wells could be observed more easily than in the three
dimensional case. On the other hand, since in near-threshold two-dimensional
systems the
wave function rapidly spreads over very large regions, a delicate trade-off
between the localization and delocalization could take place. 

Other possible extension of our problem to be studied, is its appearance in
electrostatic quantum dots \cite{Bednarek2003}.

\acknowledgments
We would like to acknowledge  SECYT-UNC,  CONICET and MinCyT C\'ordoba
for partial financial support of this project.

\end{document}